\documentstyle[12pt]{article}

\renewcommand{\ref}[1]{\raisebox{.6ex}{[#1]}}

\newcommand{\be}{\begin{equation}}
\newcommand{\ee}{\end{equation}}

\begin{document}

\title{ Implications of adiabatic phases for a vortex in a superconductor film
}

\author{ Ping Ao$^{1)}$, Qian Niu$^{2)}$, and David J. Thouless$^{1)}$
\\ {\ } \\
$^{1)}$Department of Physics, FM-15
University of Washington, Seattle, WA 98195, USA \\  {\ }  \\
$^{2)}$Department of Physics, University of Texas, Austin, TX 78712, USA   }


\begin{abstract}
Based on ideas of off-diagonal long range order and two-fluid model,
we demonstrate that adiabatic phases for a slow motion of a vortex in
a superconductor film give rise naturally to the Magnus force at finite
temperatures.
\end{abstract}


\maketitle

There is an explosive renewed interest in vortex dynamics
in superconductors\cite{super}.
A consistent and clear understanding of the basic physics for
vortex dynamics is needed.
It has been shown recently by the aid of the many-body wavefunction
at zero temperature that adiabatic phases for a slow motion of vortex line in
a superconductor give rise naturally to the Magnus force\cite{ao}.
Here we demonstrate that the same result can be extended to finite
temperatures, with the many-body wavefunction replaced by the density matrix.
For a clear representation, we consider a homogeneous 2-d
superconductor film taken to be the $x-y$ plane.
As in Ref.\cite{ao}, we first discuss the case that the magnetic field
effect is negligible.

The density matrix for N electrons is
$
   \rho_{N} = \rho_{N}( {\bf r}_{1} ... {\bf r}_{N} ;
          {\bf r}'_{1} ... {\bf r}'_{N} )
$
with $\{ {\bf r}_{j} \}$ the position vectors and spins
of the N electrons.
The one particle reduced density matrix is defined as
\be
   \rho_{1}({\bf r}; {\bf r}' ) = N
     \int\prod_{j=2}^{N} d^{2}{\bf r}_{j}
         \rho_{N}({\bf r} ... {\bf r}_{N} ;
         {\bf r}',  {\bf r}_{2} ... {\bf r}_{N} ) .
\ee
For a superconducting state, a two-fluid model may be applied for
the one particle density matrix $\rho_{1}$\cite{gorter}.
The essence of the two-fluid model is that, at any instant there exist
two real functions,
$\rho_{n}({\bf r}, {\bf r}')$ and $\rho_{s}( {\bf r}, {\bf r}' )$
such that\cite{penrose}
\[
   \rho_{1}({\bf r}; {\bf r}') = \rho_{n}({\bf r};{\bf r}')
        \exp\{ i  [ {\bf r}\cdot{\bf k}_{n}({\bf r})
       -  {\bf r}'\cdot{\bf k}_{n}({\bf r}')] \}
\]
\be
   + \rho_{s}({\bf r};{\bf r}') \exp\{i [ \phi({\bf r})
     - \phi({\bf r}') ] \}\; .
\ee
Here real functions
$\rho_{n}$, $\rho_{s}$, ${\bf k}_{n}$, and $\phi$ are the functions of time and
temperature. The total electron number density is
$ \rho({\bf r}) = \rho_{n}({\bf r}; {\bf r}) +  \rho_{s}({\bf r};{\bf r}) .$
If $\rho_{s}=0$, this would
describe the system with classical velocity field ${\bf v}_{n} =
\hbar{\bf k}_{n}/m$ with $m$ the effective mass of an electron,
and the system is in the normal state, where $\rho_{n} \rightarrow 0$
exponentially as $|{\bf r} - {\bf r}'| \rightarrow \infty $.
If $\rho_{n} = 0$, the system is at zero temperature as discussed in
Ref.\cite{ao}. We have assumed that the wavevectors ${\bf k}_{n}$
and $\nabla_{\bf r} \phi({\bf r})$ do not alter
appreciably over a distance of the order of the coherence length $\xi$.
The existence of finite superfluid density $\rho_{s}$ at finite temperatures
is the manifestation of off-diagonal long range order,
but we cannot write $\rho_{s}$ as the product of wavefunction\cite{yang}.
Instead, it should be determined from the pair wavefunction in two particle
reduced density matrix because electrons form pairs.

The Berry phase for a pure state is\cite{berry},
\be
   \Theta = i \oint_{\Gamma} d{\bf r}_{0} \cdot < \Psi_{v}({\bf r}_{0} )|
            \nabla_{{\bf r}_{0}} \Psi_{v}({\bf r}_{0}) > .
\ee
Writing this equation in a more explicit and suggestive form, the Berry phase
becomes
\[
   \Theta = i \oint_{\Gamma} d{\bf r}_{0} \cdot
           \lim_{ {\bf r}'_{0}\rightarrow {\bf r}_{0}  }
           \nabla_{{\bf r}_{0} }
           \int \prod_{j=1}^{N}d^{2}{\bf r}_{j}
\]
\be
    \rho_{N,v}(\{{\bf r}_{j}\}, {\bf r}_{0}; \{{\bf r}_{j}\}, {\bf r}'_{0} ) .
\ee
Here $\rho_{N,v}$ is the density matrix
defined from the many-body wavefunction for a vortex $\Psi_{v}$ as
$  \rho_{N,v} = \Psi_{v}(\{ {\bf r} \},{\bf r}_{0})
                \Psi_{v}^{\ast}(\{ {\bf r}' \},{\bf r}'_{0}) .$
The vortex position is ${\bf r}_{0}$.
We do not know the exact form of the Hamiltonian for vortex dynamics.
Instead, we have a good
knowledge of the wavefunction (or the density matrix)
describing a vortex state, which is the starting point to derive the Magnus
force by calculating the adiabatic phases.
This method is essentially the
Born-Oppenheimer approximation where the vortex coordinates are slow ones,
and has been put into a more general perspective with the aid of the Berry
phase\cite{wilczek}.
Eq.(4) is the generalization of the Berry phase from a pure state
to a mixed state.

In the region far away from the vortex core where the supercurrent generated
by the vortex is small, the two-fluid model can be applied.
The ansatz for the one particle reduced density matrix
of the vortex state at a finite temperature can be written as eq.(2), with
\be
   \phi({\bf r}) =   \theta({\bf r}-{\bf r}_{0} )
              + \frac{m}{\hbar} {\bf v}_{s}\cdot {\bf r} \; ,
\ee
where  $\theta = q_{v} \arctan[(y-y_{0})/(x-x_{0})]/2 $
with $q_{v}$ the winding number of a vortex and ${\bf v}_{s}$ is the
superfluid velocity.

Using eqs.(4), (1), (2), and (5), the Berry phase for a vortex adiabatically
moving around an arbitrary close loop $\Gamma$ is
\be
   \Theta = - \oint_{\Gamma} d{\bf r}_{0} \int d^{2}{\bf r}
           \nabla_{{\bf r}_{0} }  \theta({\bf r}-{\bf r}_{0} )
           \frac{\rho_{s,v}({\bf r}, {\bf r}_{0} ) }{2} ,
\ee
where there is no contribution from the normal fluid $\rho_{n}$.
Ignoring the contribution from the core as done in Ref.\cite{ao},
we obtain the Berry phase
\be
   \Theta = - 2\pi q_{v} \frac{\rho_{s} }{2} S(\Gamma) \; .
\ee

The dynamical phase corresponds to the potential
energy of the system at a finite temperature.
For given density matrix, the potential energy of the system is given by
$U({\bf r}_{0}) = \int \prod_{j=1}^{N} d{\bf r}_{j}
 \lim_{ {\bf r}_{j}' \rightarrow {\bf r}_{j} }
  H(\{{\bf r}_{j}\}) \rho_{N,v}(\{{\bf r}_{j}\},{\bf r}_{0};
  {\bf r}_{j}', {\bf r}_{0} ) $.
One and two particle reduced matrices are involved for
the usual BCS Hamiltonian $H$.
The one particle reduced density matrix is given by eq.(2).
For the two particle reduced density matrix
it may be reasonable to assume that this density matrix is
close to the one without the vortex, except near the vortex core where the
contribution can be put into the core energy.
Then the potential energy depending on the position of the vortex
in the presence of an external current is
\be
   U( {\bf r}_{0} ) = -
      \frac{\rho_{s}}{2} \hbar q_{v} {\bf r}_{0} \cdot
        {\bf v}_{s} \times \hat{z}  \; .
\ee
There is no dependence on the normal current.

Combining eqs.(7) and (8), we obtain the Magnus force at a finite temperature:
\be
   {\bf F}_{m} = \frac{\rho_{s}}{2} \hbar q_{v} ({\bf v}_{s}
                   - \dot{\bf r}_{0} )\times \hat{z}  \; ,
\ee
which is the same as eq.(9) in Ref.\cite{ao}, except
that here the superfluid electron number density is temperature dependent.

As demonstrated in Ref.\cite{ao},
the electromagnetic field associated with a vortex
does not affect the Magnus force
because of the charge neutrality of the superconducting film.
The disorder does not affect its existence either,
except for a possible reduction of the
superfluid density due to the localization of some Cooper pairs
and a pinning force due to the introduced inhomogeneity.

Finally, we wish to point out that neither the Iordanskii
force\cite{iordanskii} nor the conservative force recently discussed by
Ferrell\cite{ferrell} seem to appear in the present approach.

We thank M.Y. Choi and Y. Tan for useful discussions.
This work was supported by US National Science Foundation under Grant No. DMR
89-16052.


\begin{thebibliography}{99}
\bibitem{super}
  N. Giordano, Phys. Rev. Lett. {\bf 61}, 2137 (1988);
  T.W. Jing and N.P. Ong, Phys. Rev. {\bf B42}, 10781 (1990);
  S.J. Hagen {\it et al}., Phys. Rev. {\bf B43}, 6246 (1991);
  M.P.A. Fisher {\it et al}., Phys. Rev. Lett. {\bf 66}, 2931 (1991);
  Y. Liu {\it et al}., {\it ibid}, {\bf 68}, 2224 (1992);
  P. Ao, {\it ibid}, {\bf 69}, 2997 (1992);
  D.A. Huse {\it et al}., Nature {\bf 358}, 553 (1992);
  P. Ao, J. Low Temp. Phys. {\bf 89}, 543 (1992);
  A.C. Mota {\it et al}., Phys. Scr. {\bf T45}, 69 (1992);
  R. Fazio {\it et al}., Helv. Phys. Acta {\bf 65}, 228 (1992).
\bibitem{ao}
  P. Ao and D.J. Thouless, Phys. Rev. Lett. {\bf 70}, 2158 (1993).
\bibitem{gorter}
  C.J. Gorter and H. Casimir, Physica {\bf 1}, 306 (1934).
\bibitem{penrose}
  O. Penrose, Phil. Mag. {\bf 42}, 1373 (1951).
\bibitem{yang}
  C.N. Yang, Rev. Mod. Phys. {\bf 34}, 694 (1962).
\bibitem{berry}
 M.V. Berry, Proc. Roy. Soc. {\bf A392}, 45 (1984).
\bibitem{wilczek}
 {\it Geometric Phases in Physics}, edited by A. Shapere and F.
 Wilczek, World Scientific, Singapore, 1989.
\bibitem{iordanskii}
  S.V. Iordanskii, Ann Phys. {\bf 29}, 335 (1964).
\bibitem{ferrell}
  R.A. Ferrell, Phys. Rev. Lett. {\bf 68}, 2524 (1992).
\end{thebibliography}
\end{document}